\documentstyle[preprint,epsf,amssymb,aps]{revtex}
\draft
\tighten

\newcommand{\be}[1]{
\begin{eqnarray}\label{#1}}
\newcommand{\ee}{\end{eqnarray}}
\newcommand{\ci}[1]{\cite{#1}}
\newcommand{\re}[1]{(\ref{#1})}

\newcommand{\spin}[1]{ \langle\mskip-3mu \langle{#1}
\rangle\mskip-3mu\rangle}
\begin{document}
\renewcommand{\thefootnote}{\fnsymbol{footnote}}

\begin{flushright}
\begin{tabular}{l}
TPR-00-22 \\ RUB-TP2-23/00 \\ MKPH-T-00-25 \\
\end{tabular}
\end{flushright}

\begin{center}
{\bf\Large DVCS on the nucleon : study of the twist-3 effects}

\vspace{0.5cm} N. Kivel$^{a,b}$, M.V. Polyakov$^{b,c}$, M. Vanderhaeghen$^d$

\begin{center}
{\em $^a$ Institut f\"ur Theoretische Physik, Universit\"at Regensburg,
D-93040 Regensburg, Germany \\
$^b$ Petersburg Nuclear Physics Institute, 188350, Gatchina, Russia \\
$^c$ Institut f\"ur Theoretische Physik II,Ruhr-Universit\"at Bochum,
D-44780 Bochum, Germany \\
$^d$ Institut f\"ur Kernphysik, Johannes Gutenberg-Universit\"at,
D-55099 Mainz, Germany}

\end{center}
\end{center}




\begin{abstract}
We estimate the size of the twist-3 effects
on deeply virtual Compton scattering (DVCS) observables,
in the Wandzura-Wilczek approximation.
We present results in the valence region
for the DVCS cross sections, charge asymmetries and single spin
asymmetries, to twist-3 accuracy.

\end{abstract}

\section{\normalsize \bf Introduction}

Deeply virtual Compton scattering (DVCS) \cite{DVCS1,DVCS2,DVCS3} has been
studied intensively in recent years as a new hard probe to access
generalized parton distributions of the nucleon, the so-called
skewed parton distributions (SPDs).
\newline
\indent
After the first experimental evidence for DVCS on the proton 
\cite{ZEUS,H1,Amarian,JLAB}, it is the right time to address the
problem of the extraction of the twist-2 SPDs from DVCS observables.
In order to be able to extract the twist-2 SPDs from DVCS
observables at accessible values of the `hard' scale $Q$, one first needs to
estimate the effects of higher twist (power suppressed)
contributions to those observables. The first power correction to
the DVCS amplitude is of order $O(1/Q)$, hence it is called
twist-3. The twist-3 corrections to the DVCS amplitude have been
derived recently by several groups
\cite{Anikin,Penttinen,BM,RW} using different approaches.
It was shown that the twist-3 part of the amplitude depends on a new
type of SPDs which can be related to the twist-2 SPDs with the help of
Wandzura-Wilczek (WW) relations. These relations are based on an
assumption that the nucleon matrix elements of mixed quark-gluon
operators are small relative to matrix elements of symmetric quark
operators, for details see \cite{BM,KPST,KP,RW}. Originally, the WW
relations were derived for the polarized structure functions
$g_1(x)$ and $g_T(x)$ \cite{WW}. Recent experimental measurements
\cite{g2izmer} of these polarized structure functions indicate that
the WW relations are satisfied to good accuracy. In the theory of the instanton
vacuum, the smallness of the corrections to the WW relations for
$g_T(x)$ and $h_L(x)$ can be related to smallness of the packing
fraction of the instantons in the QCD vacuum, for details see
\cite{BPW,DP}.
\newline
\indent
Our aim here is to study the twist-3 effects on DVCS observables
for a nucleon target, in the Wandzura-Wilczek approximation
\footnote{Recently the twist-3 corrections
to the DVCS amplitude on the pion target were studied in Ref.~\cite{BMKS}}.
Although on the one hand,
the twist-3 effects are formally a correction when one aims
to extract twist-2 SPDs from DVCS observables,
they provide on the other hand an additional handle in the extraction
of those same twist-2 SPDs.
Indeed, in the WW approximation the twist-3 SPDs are fixed completely in
terms of twist-2 SPDs and their derivatives.
\newline
\indent
In this work, we restrict our analysis to the leading order in $\alpha_s(Q^2)$,
because the NLO corrections were computed only for the twist-2 amplitude.
Since it was shown that the NLO correction to the
twist-2 DVCS amplitude can be sizeable \cite{JiOs,MaPi,BeMu}, the NLO
analysis for the twist-3 amplitude remains to be investigated.
\newline
\indent
In the present work, we firstly give in section \ref{sec2} the
expressions for the DVCS amplitude on the nucleon to twist-3 accuracy,
as calculated in the WW approximation.
\newline
\indent
In section \ref{sec3}, we describe the modelization of the SPDs that
will be used to provide estimates for DVCS observables. We propose a
two-component parametrization for the SPDs, so as to satisfy the
polynomiality conditions.
\newline
\indent
In section \ref{sec4}, we show our results for DVCS observables, and
discuss the importance of the twist-3 effects on the DVCS
cross sections, charge asymmetries and single spin asymmetries in the
valence region.
\newline
\indent
Finally, we give our conclusions in section \ref{sec5}.

\section{\normalsize \bf DVCS amplitude on the nucleon at the twist-3
accuracy in Wandzura-Wilczek approximation}
\label{sec2}

The amplitude of the virtual Compton scattering process
\be{proc}
\gamma^*(q)+N(p)\to \gamma(q') +N(p') \, ,
\ee
is defined in
terms of the nucleon matrix element of the $T$-product of two
electromagnetic currents~:
\be{T:def}
T^{\mu\nu}=i\int d^4x\ e^{-i (q\cdot x)}\langle p'|T\left[J_{\rm
e.m.}^\mu (x) J_{\rm e.m.}^\nu(0)\right]|p\rangle\, ,
\ee
where the four-vector index $\mu$ ($\nu$) refers to the virtual (real)
photon.
\newline
\indent
In the Bjorken limit, where $-q^2=Q^2\to\infty$, $2(p\cdot q)\to \infty$,
with $x_B = Q^2 / 2(p\cdot q)$ constant, and $t \equiv (p-p')^2\ll Q^2$,
the DVCS amplitude on the nucleon to the order $O(1/Q)$
has the form \cite{Penttinen,BM}~:
\be{T} T^{ \mu
\nu}&=& \frac12 \int_{-1}^1 dx\quad \biggl\{ \left[(-g^{\mu
\nu})_\perp-\frac{P^\nu\Delta_\perp^\mu}{(P \cdot q')}  \right] n^\beta
F_\beta (x,\xi) C^+(x,\xi) \nonumber \\[4mm]&-& \left[(-g^{\nu
k})_\perp-\frac{P^\nu\Delta_\perp^k}{(P \cdot q')}  \right]
i\epsilon^{\perp\mu}_{k } n^\beta \widetilde F_\beta (x,\xi)
C^-(x,\xi) \nonumber \\[4mm]&-& \frac{(q+4\xi P)^\mu}{(P \cdot q)}
\left[(-g^{\nu k})_\perp-\frac{P^\nu\Delta_\perp^k}{(P \cdot q')}  \right]
\left\{ F_k(x,\xi)C^+(x,\xi)- i\epsilon^\perp_{k \rho}\widetilde
F^\rho (x,\xi) C^-(x,\xi)\right\} \biggr\} \, ,
\ee
where to the twist-3 accuracy: \be{lce} P&=&\frac12(p+p')=n^*, \quad
\Delta = p'-p =-2\xi P+\Delta_\perp, \, \nonumber \\[4mm]
q&=&-2\xi P+\frac{Q^2}{4\xi}n, \quad
q'=q-\Delta=\frac{Q^2}{4\xi}n-\Delta_\perp \ee where $p, p'$ are
the momenta of the initial and final nucleon and $q,q'$ are the momenta of
the initial and final photon respectively. The light-like vectors
$n$ and $n^*$ are normalized as $(n\cdot n^*)=1$. We also
introduce the metric and totally antisymmetric tensors in the two
dimensional transverse plane ($\varepsilon_{0123}=+1$)~:
\be{gt}
(-g^{\mu \nu})_\perp = -g^{\mu \nu}+ n^\mu n^{* \nu}+n^\nu n^{* \mu},
\quad \epsilon^\perp_{\mu \nu}=
\epsilon_{\mu \nu \alpha\beta}n^\alpha n^{* \beta} \, .
\ee
The leading order coefficient functions are: \be{alf} \nonumber
C^\pm(x,\xi)=\frac{1}{x-\xi+i\varepsilon}\pm
\frac{1}{x+\xi-i\varepsilon}.
\ee
\indent
In the expression for the DVCS amplitude to the twist-3 accuracy
the first two terms correspond to the scattering of transversely
polarized virtual photons. This part of the amplitude depends
only on twist-2 SPDs $H,E$ and $\widetilde H, \widetilde E$ and was
anticipated in refs.~\cite{GuichonVander,VGG}.
The third term in Eq.~\re{T} corresponds to
the contribution of the longitudinal polarization of the virtual
photon. This term depends only
on new `transverse' SPDs $F_\perp^\mu$ and $\widetilde F_\perp^\mu$. Defining
the polarization vector of the virtual photon as
\be{epsL}
\varepsilon_L^\mu(q)=\frac{1}{Q}\biggl(
2\xi P^\mu+\frac{Q}{4\xi} n^\mu
\biggr)\, ,
\ee
we can easily calculate the DVCS amplitude for longitudinal
polarization of the virtual photon ($L\to T$ transition),
which is purely of twist-3~:
\be{LtoT}
\varepsilon_\mu^L \, T^{\mu\nu}=\frac{2\xi}{Q}\int_{-1}^1dx
\ \Biggl(F_\perp^\nu\  C^+(x,\xi)- i\varepsilon_\perp^{\nu k}
\widetilde F_{\perp k}\  C^-(x,\xi)
\Biggr)\, .
\ee
The skewed parton distributions
$F_\mu$ and $\widetilde F_\mu$ can be related to the twist-2 SPD's
$H,E,\widetilde H$ and $\widetilde E$ with help of
Wandzura-Wilczek relations. To derive these relations one assumes
that the non-forward nucleon matrix elements of gauge invariant operators
of the type $\bar \psi G \psi$ are small. The WW relations for the
case of nucleon SPDs have the form
\cite{BM,KP}~:
\be{F}
F^{WW}_\mu(x,\xi)&=&  \frac{\Delta_\mu}{2\xi}\spin{\frac
1 M}E(x,\xi)- \frac{\Delta_\mu}{2\xi}\spin{\gamma_+}(H+E)(x,\xi)+
\nonumber\\&& +\int_{-1}^{1}du\
G_\mu(u,\xi)W_{+}(x,u,\xi)+i\epsilon_{\perp \mu k}
\int_{-1}^{1}du\  \widetilde G^k (u,\xi)W_{-}(x,u,\xi)\, , \\
[4mm] \widetilde F^{WW}_\mu(x,\xi)&=&
\Delta_\mu\frac12 \spin{\frac{\gamma_5}{M}}\widetilde E(x,\xi)-
\frac{\Delta_\mu}{2\xi}\spin{\gamma_+\gamma_5}\widetilde H(x,\xi)+
\nonumber\\ && +\int_{-1}^{1}du\ \widetilde
G_\mu(u,\xi)W_{+}(x,u,\xi)+i \epsilon_{\perp \mu k}
\int_{-1}^{1}du\  G^k (u,\xi)W_{-}(x,u,\xi) \, .
\ee
The following notations are used~:
\be{G}
G^\mu(u,\xi)&=& \spin{\gamma^\mu_\perp}(H+E)(u,\xi)+
\frac{\Delta_\perp^\mu}{2\xi} \spin{\frac 1
M}\biggl[u\frac{\partial}{\partial u}+ \xi\frac{\partial}{\partial
\xi} \biggl] E(u,\xi)- \nonumber \\[4mm]&&
-\frac{\Delta_\perp^\mu}{2\xi}
\spin{\gamma_+}\biggl[u\frac{\partial}{\partial u}+
\xi\frac{\partial}{\partial \xi}\biggl] (H+E)(u,\xi) \, ,
\ee
\be{tG}
\widetilde G^\mu (u,\xi)& =&\spin{\gamma^\mu_\perp\gamma_5} \widetilde H(u,\xi)
+\frac12\Delta_\perp^\mu \spin{\frac{\gamma_5}{M}}
\biggl[1+u\frac{\partial}{\partial u}+\xi\frac{\partial}{\partial
\xi}\biggl] \widetilde E(u,\xi)- \nonumber\\[4mm]&&
-\frac{\Delta_\perp^\mu}{2\xi}\spin{\gamma_+\gamma_5}
\biggl[u\frac{\partial}{\partial u}+\xi\frac{\partial}{\partial
\xi}\biggl] \widetilde H(u,\xi) \, .
\ee
\indent
The functions $W_{\pm}(x,u,\xi)$ are the so-called
Wandzura-Wilczek  kernels introduced in Ref.~\cite{KP}. They are
defined as~:
\be{Wpm}
W_{\pm}(x,u,\xi)&=& \frac12\biggl\{
\theta(x>\xi)\frac{\theta(u>x)}{u-\xi}-
\theta(x<\xi)\frac{\theta(u<x)}{u-\xi} \biggl\} \nonumber
\\[4mm]&&\mskip-10mu \pm\frac12\biggl\{
\theta(x>-\xi)\frac{\theta(u>x)}{u+\xi}-
\theta(x<-\xi)\frac{\theta(u<x)}{u+\xi} \biggl\}.
\ee
The sandwiching between nucleon Dirac spinors is denoted by
$\spin{\ldots}=\bar U(p')\ldots U(p)$ and $M$ denotes the nucleon mass.
The quarks flavor dependence in the amplitude can be easily restored by the
substitution~:
\be{flav}
F_\mu \, (\widetilde F_\mu) \to
\sum_{q=u,d,s, \dots}e_q^2\ F_\mu^q \, (\widetilde F_\mu^q) \, .
\ee
\indent
The amplitude \re{T} is electromagnetically gauge invariant, i.e.
\be{Ginv} q_\mu T^{\mu\nu}=(q-\Delta)_\nu T^{\mu\nu}= 0\, ,
\ee
formally to the accuracy $1/Q^2$. In order to have `absolute' transversality
of the amplitude we keep in the expression \re{T} terms of the
$\Delta^2/Q^2$ order, applying the prescription of
\cite{GuichonVander,VGG}~:
\be{transvers}
-g^{\mu\nu}_\perp \, \to \, -g^{\mu\nu}_\perp \,-\,
\frac{P^\nu\Delta_\perp^\mu}{(P \cdot q')} \, ,
\ee
for the twist-3 terms in the amplitude. Formally such terms are beyond our
accuracy and they do not form a complete set of $1/Q^2$
contributions, but we prefer to work with the DVCS amplitude,
satisfying Eq.~(\ref{transvers}) exactly.
\newline
\indent
The convolution of the leading order Wilson coefficients \re{alf} with
the WW kernels can be performed analytically
(see also \cite{BMKS}). For example the part
of the amplitude which contains WW kernels can be simplified using
the following result for the integration over $x$~:
\be{conv}
&&\int_{-1}^1 dx \; \Biggl\{\, \frac{1}{x-\xi+i\varepsilon}\ \int_{-1}^1 du\
(W_+ - W_-)\biggl[\, G^\mu -i \varepsilon_\perp^{\mu k}\widetilde G_k \,\biggr]
\nonumber \\
&&\hspace{1.3cm}\,+\,\frac{1}{x+\xi-i\varepsilon}\ \int_{-1}^1 du\
(W_+ + W_-)\biggl[\, G^\mu +i \varepsilon_\perp^{\mu k}\widetilde G_k \,\biggr]
\Biggr\} \nonumber\\
&=& -i\pi \int_{\xi}^{1}du\ \frac{G^\mu -i \varepsilon_\perp^{\mu
k}\widetilde G_k}{u+\xi}
\,-\,i\pi \int_{-1}^{-\xi}du\ \frac{G^\mu +i \varepsilon_\perp^{\mu
k}\widetilde G_k}{u-\xi}
\nonumber \\
&&+\,\int_{-1}^1 du\ \Biggl\{ \, \frac{G^\mu -i \varepsilon_\perp^{\mu
k}\widetilde G_k }{u+\xi}\ \ln \biggl|\frac{u-\xi}{2\xi} \biggr|
\,+\,\frac{G^\mu +i \varepsilon_\perp^{\mu
k}\widetilde G_k }{u-\xi}\ \ln \biggl|\frac{u+\xi}{2\xi} \biggr| \,
\Biggr\} \, .
\ee
Note that all integrals over $u$ are convergent and the
strongest (integrable) singularity of the integrand is logarithmic only.

\section{\normalsize \bf Modelization of SPDs }
\label{sec3}

\begin{figure}[h]
\epsfxsize=12cm
\centerline{\epsffile{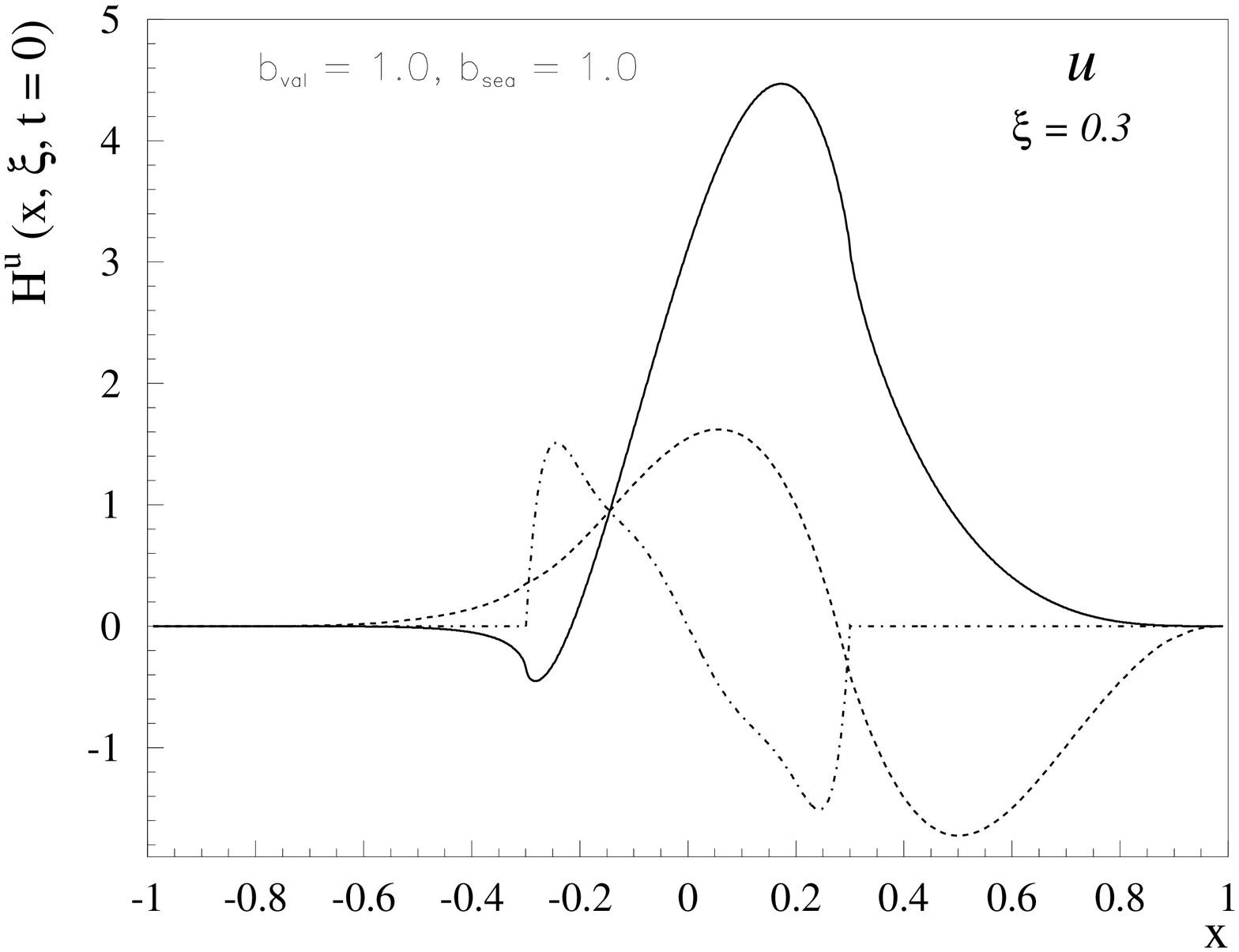}}
\caption{{\em SPD $H^u(x,\xi)$ for the u-quark 
obtained from the double distribution model \re{modelb} (solid curve);
$(1+x\partial_x+\xi\partial_\xi) H^u(x,\xi)$ (dotted curve), and D-term
\re{dterm} (dashed-dotted curve) at a fixed value of $\xi=0.3$.}}
\label{iaa}
\end{figure}

In order to interpret DVCS observables, we next discuss the
parametrization of the SPDs. As we address here hard exclusive
reactions in the valence region at relatively small values of $-t$,
we factorize the $t$-dependence of the SPDs
by the corresponding form factor, so as to satisfy the first sum rule.
We stress that the factorization of the $t$-dependence
of SPDs is not expected to hold when increasing the value of $t$. 
For example, the calculations of the SPDs in the 
chiral quark soliton model \cite{pp4} showed
that at $t$ about $-0.8$~GeV$^2$ the deviations from the
factorization ansatz for the $t$-dependence of SPDs is as large as
about 25\% in some $x$-range. 
In the calculations presented here, we limit ourselves to
relatively small values of $-t$ = 0.25 GeV$^2$ and assume the
factorization ansatz. A more realistic parametrization of SPDs 
at larger values of $t$, remains to be done in a future work. 
We next discuss the resulting $t$-independent part.
\newline
\indent
For the function $H^q$ (for each flavor $q$),
the $t$-independent part is parametrized by a two-component form~:
\be{dd}
H^q(x, \xi) \,=\, H^q_{DD}(x, \xi) \,+\,\frac{1}{N_f} \ D(\frac{x}{\xi})\, ,
\ee
where $H^q_{DD}$ is the part of the SPD which is obtained as a
one-dimensional section of a two-variable double distribution $F^q$
\cite{Raddouble}, imposing a particular skewedness $\xi$, as~:
\be{dd2}
H^q_{DD}(x,\xi)=
\int_{-1}^{1}d\beta\
\int_{-1+|\beta|}^{1-|\beta|} d\alpha\
\delta(x-\beta-\alpha\xi)\  F^q(\beta,\alpha)\ \, ,
\ee
and where the D-term contribution\footnote{The name ``D-term"
is derived from (more or less arbitrary)
notations used in Ref.~\cite{PW} and to some extent
is confusing because of its similarity to the terminology used
in supersymmetric theories, especially when the first term in Eq.~(\ref{dd})
is called ``F-term".}
$D$
completes the parametrization of SPDs, restoring the correct polynomiality
properties \cite{Jirev} of SPDs \cite{PW}.
The D-term contribution to SPDs has a support only for $|x|\leq
|\xi|$, so that it is `invisible' in the forward limit.
The D-term is an isoscalar contribution, and adds therefore the same
function for each flavor (in Eq.~(\ref{dd}), $N_f$ = 3 is the number of
 active flavors).
\newline
\indent
For the double distributions, entering Eq.~(\ref{dd2}),
we use the following model suggested
in \cite{Raddouble}~:
\be{modelb}
F^q(\beta,\alpha)=
 \frac{\Gamma(2b+2)}{2^{2b+1}\Gamma^2(b+1)}\
\frac{\bigl[(1-|\beta|)^2-\alpha^2\bigr]^{b}}{(1-|\beta|)^{2b+1}}\
q(\beta)\, ,
\ee
with $q(\beta)$ is the corresponding forward parton distribution ensuring
the correct forward limit for the SPD $H(x,0)=q(x)$. We use the
phenomenological forward parton distributions (including their
evolution) as input.
In Eq.~(\ref{modelb}),
the parameter $b$ characterizes the strength of the $\xi$
dependence of the SPD $H(x,\xi)$, the limiting case $b\to\infty$
corresponds to the $\xi$ independent SPD $H(x,\xi)=q(x)$.
The power $b$ in eq.~(\ref{modelb}) is a free parameter
for the valence contribution ($b_{val}$)
and for the sea/antiquark contribution ($b_{sea}$) to the SPD,
which can be used as fit parameter in this approach
to extract SPDs from DVCS observables.
The twist-2 DVCS predictions of Ref.~\cite{VGG2}, correspond to the choice :
$b_{val} = b_{sea} = 1.0$.
In the following, we also show all predictions for the value~:
$b_{val} = b_{sea} = 1.0$, in order to have a point of comparison, and
refer to a future work for a more systematic study of the dependence of DVCS
observables on the shape of the profile function.
\newline
\indent
The D-term contribution to the singlet SPDs $H(x,\xi)$ and
$E(x,\xi)$ has the following form~:
\begin{eqnarray}
H^{\rm D-term}(x,\xi)&=&\theta(\xi-|x|)\ D\left(\frac{x}{\xi}
\right)\, ,
\label{dterm:contrib} \\
E^{\rm D-term}(x,\xi)&=&-\theta(\xi-|x|)\ D\left(\frac{x}{\xi}
\right)\, \label{dterm:contrib2} .
\end{eqnarray}
Note that the D-term contribution is cancelled in the combination
$H+E$ to ensure the polynomiality condition (for odd $N$)~:
\be{pol}
\int_{-1}^1dx\ x^N \left( H(x,\xi)+E(x,\xi)\right)={\rm polynomial\
of\ the\ order\ }(N - 1) \, {\rm in\ }\xi\, .
\ee
Note that the $N$-th Mellin moments of $H$ and $E$ separately are
polynomials of order $N+1$ in $\xi$, only in the combination
(\ref{pol}) the highest powers are cancelled.
As model for the SPD $E(x,\xi)$, we shall only use the
contribution of the D-term \re{dterm:contrib2}. More detailed
studies of DVCS observables with the function $E(x,\xi)$ computed
in the chiral quark-soliton model \ci{Pet98,PetB98} will be
published elsewhere.
\newline
\indent
The function $D(\alpha)$ can be expanded in odd Gegenbauer polynomials :
\be{dterm}
D(z,t=0)= (1-z^2) \biggl[ d_1 \ C_1^{3/2}(z)
+ d_3\ C_3^{3/2}(z) + d_5 \ C_5^{3/2}(z) + ...\biggr] \, .
\ee
For the moments $d_1, d_3, d_5$, we use the estimate
which is based on the calculation of SPDs
in the chiral quark soliton model \cite{Pet98}
at a low normalization point $\mu\approx 0.6$~GeV, 
which gives~:
\begin{equation}
d_1 \approx -4.0, \hspace{1cm} d_3 \approx -1.2, \hspace{1cm} d_5 \approx -0.4,
\label{eq:dterm_exp}
\end{equation}
and higher moments (denoted by the ellipses in Eq.~(\ref{dterm}))
are small and are neglected in the following.
\footnote{In what follows we neglect the scale dependence of the
  result (\ref{eq:dterm_exp})
as the uncertainties in modeling of the D-term are larger than the
logarithmic scale dependence, for the scales considered in our calculations.}
Notice the negative sign of the Gegenbauer coefficients for
the D-term in Eq.~(\ref{eq:dterm_exp}), as obtained in the chiral
quark-soliton model.
\newline
\indent
It is interesting to note that for the case of SPDs in the pion, the value of
the coefficient $d_1$ in the parametrization of
the D-term (\ref{dterm}) can be computed in a model independent way.
To do this, we can use the soft-pion theorem for the singlet SPD
in the pion derived in \cite{MVP98}. This soft-pion  theorem states
that the singlet SPD in the pion vanishes for $\xi = \pm 1$~:
\be{soft-pion}
\sum_q H_q^{(\pi)}(x,\xi= \pm 1,t=0)=0\, .
\ee
Hence (see also \cite{PW}),
\be{hpion}
\int_{-1}^1 dx \, x \, \left(\sum_q H_q^{(\pi)}(x,\xi,t=0) \right)
\, =\, \left(1 - \xi^2 \right) \, M_2^Q \, .
\ee
Evaluating Eq.~(\ref{hpion}) at $\xi = 0$, determines~:
\be{m2}
M_2^Q= \int_0^1dx \, x \, \sum_q [q^{(\pi)}(x) + \bar q^{(\pi)}(x)]\, ,
\ee
being the fraction of the momentum carried by the quarks and antiquarks in the
pion.
As the highest power in $\xi$ in Eq.~(\ref{hpion}) (i.e. the term in
$\xi^2$) originates solely from the D-term, one easily obtains the
following expression for pion D-term~:
\be{pionDterm}
D^{(\pi)}(z)=-\frac{5 M_2^Q} 4\ (1-z^2)
\biggl[\ C_1^{3/2}(z)
+ ...\biggr]\, .
\ee
We see therefore that the first Gegenbauer coefficient of the pion D-term
is negative and strictly nonzero.
The coincidence of the signs in the nucleon and pion
D-terms hints that the D-term in the nucleon is intimately related
to the spontaneous breaking of the chiral symmetry.
\newline
\indent
When using the parametrization of Eq.~(\ref{dd}) for the pion,
the soft pion theorem (\ref{soft-pion}) imposes a condition between
both terms, and fixes the pion D-term completely in terms of
the isosinglet double distribution as follows~:
\be{pionDF}
D^{(\pi)}(z)=-
\int_{-1}^{1}d\beta\
\int_{-1+|\beta|}^{1-|\beta|} d\alpha\
\delta(z-\beta-\alpha)\  \sum_q F^q(\beta,\alpha)\ \, .
\ee
Therefore, when studying DVCS off the pion it is enough to model
the double distribution, the corresponding D-term is restored by
using the soft pion theorems. For the nucleon, both terms in
Eq.~(\ref{dd}) are parametrized independently.
\newline
\indent
For the calculation of the twist-3 terms, we need derivatives of SPDs
$H(x,\xi)$ which can be computed in terms of derivatives of double
distributions, using the obvious result~:
\be{deriv}
\biggl(x\frac{\partial}{\partial x}+\xi\frac{\partial}{\partial \xi}\biggr)
H_{DD}(x,\xi)=-H_{DD}(x,\xi)+
\int_{-1}^{1}d\beta\ \frac{\beta}{\xi}\
\int_{-1+|\beta|}^{1-|\beta|} d\alpha\
\delta(x-\beta-\alpha\xi)\  \frac{\partial
  F(\beta,\alpha)}{\partial\alpha} \, .
\ee
For the model of Eq.~(\ref{modelb}),
the derivative with respect to $\alpha$ in Eq.~(\ref{deriv})
is easily performed:
\be{derivmodel}
\frac{\partial F(\beta,\alpha)}{\partial\alpha}=
-\frac{b\alpha}{2^{2b}}\ \frac{\Gamma(2b+2)}{\Gamma^2(b+1)}\
\frac{\bigl[(1-|\beta|)^2-\alpha^2\bigr]^{b-1}}{(1-|\beta|)^{2b+1}}\
q(\beta)\, .
\ee
Note that although generically the derivatives of a SPD $H(x,\xi)$
with respect to $x$
and $\xi$ are discontinuous at the points $x=\pm\xi$, the
combination $(x\partial_x+\xi\partial_\xi) H_{DD}(x,\xi)$ is continuous
in these points.
The contribution of the D-term to the `transverse' SPDs
$F_\perp^\mu$ and $\widetilde F_\perp^\mu$ has been computed in
Ref.~\cite{KP} with the result~:
\be{Dterm3} \nonumber F_\mu^{\rm WW\
D-term}(x,\xi)&=&-\frac{\Delta_\mu}{2\xi\ M}\spin{1}
D\left(\frac{x}{\xi}\right)\ \theta(|x|\le\xi)\, ,\\ \widetilde
F_\mu^{\rm WW\ D-term}(x,\xi)&=&0\, . \ee
Therefore the contribution of the D-term to the DVCS amplitude
can be easily added separately from the contribution of the
double distribution part $H_{DD}$ of Eq.~(\ref{modelb}). Obviously the
D-term contributes to the real part of the amplitude only.
\newline
\indent
In Fig.~\ref{iaa}, we plot for illustrative purposes
$H^u(x,\xi)$ obtained from the model \re{modelb}
with $b_{\rm val}=b_{\rm sea}=1$,
using the MRST98 parametrization \cite{mrst98} for the
forward quark distribution as input.
We also show separately the contribution of the
D-term \re{dterm} to this function, as well as the
combination $(1+x\partial_x+\xi\partial_\xi) H^u(x,\xi)$ which
enters the twist-3 SPDs in WW approximation.
\newline
\indent
To model the SPD $\widetilde H$, we make use of its representation in
terms of double distributions in the same way as for the function
$H$, see Eqs.~(\ref{dd},\ref{dd2},\ref{modelb}). For the forward polarized
parton distributions, we use (as in \cite{VGG2})
the parametrization of Ref.~\cite{Leader98}.
Note that calculations in the
chiral quark-soliton model of the functions $\widetilde H$\cite{pp4} show that
the corresponding D-term is very small and can be neglected.
\newline
\indent
As model for $\widetilde E$ we take the contribution of the pion
pole \cite{pionpole,pp2,pp3,pp4} of the form:
\be{pipo}
\widetilde E^{\rm pion\ pole}(x,\xi)=\frac{4 g_A^2
M^2}{-t+m_\pi^2}\ \frac 1\xi \varphi_\pi\left(\frac x\xi
\right)\theta(|x|\le\xi)\, ,
\ee
where $g_A$ is the axial charge of the nucleon and $\varphi_\pi(z)$
is the pion distribution amplitude (DA).
The calculations are performed with an asymptotic DA for the pion.
Again the contribution of the pion pole to the
`transverse' functions $F_\perp^\mu$ and $\widetilde F_\perp^\mu$
has been computed in \cite{KP} with the result~:
\be{pipo3} \nonumber F_\mu^{WW,{\rm pion\ pole}}(x,\xi)&=&0\, ,\\
\widetilde F_\mu^{WW,{\rm pion\ pole}}(x,\xi)&=&
\frac{\Delta_\mu}{2M}\spin{\gamma_5}\frac{4 g_A^2
M^2}{-t+m_\pi^2}\ \frac 1\xi \varphi_\pi\left(\frac x\xi
\right)\theta(|x|\le\xi)\, .
\ee

\section{\normalsize \bf Results and discussion}
\label{sec4}

\begin{figure}[h]
\epsfxsize=13cm
\centerline{\epsffile{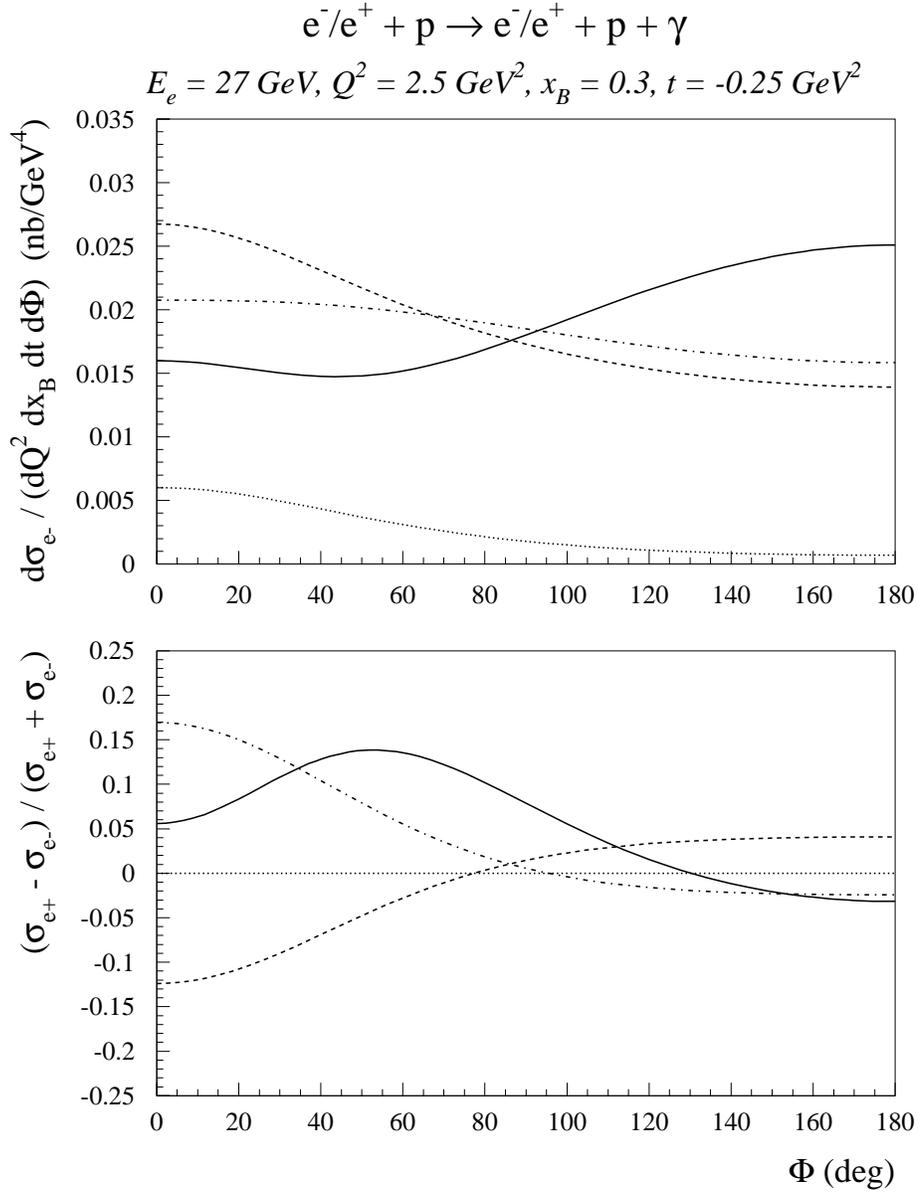}}
\caption{{\em Invariant cross section for the
$e^- p \to e^- p \gamma$ reaction (upper panel) and
DVCS charge asymmetry (lower panel) at $E_e$ = 27 GeV, for the DVCS
kinematics as indicated in the figure.
Dotted curves : BH
contribution; dashed curves : BH + twist-2 DVCS (without
D-term); dashed-dotted curves : BH + twist-2 DVCS (with D-term);
full curves : BH + twist-3 DVCS. }}
\label{fig:cross1}
\end{figure}

\begin{figure}[h]
\epsfxsize=13cm
\centerline{\epsffile{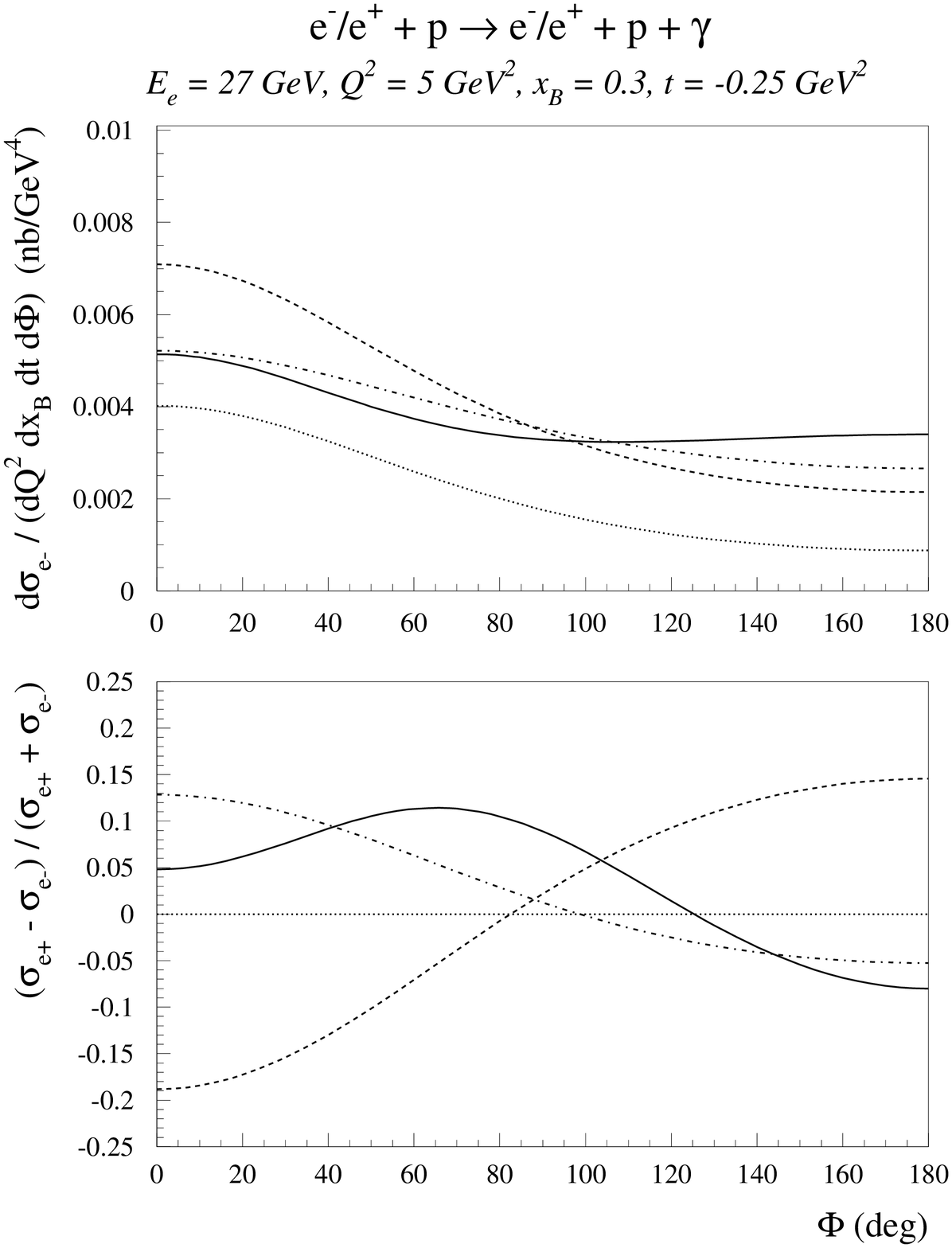}}
\caption{{\em Same as Fig.~\ref{fig:cross1}, but for $Q^2$ = 5 GeV$^2$.}}
\label{fig:cross2}
\end{figure}

\begin{figure}[h]
\epsfxsize=13cm
\centerline{\epsffile{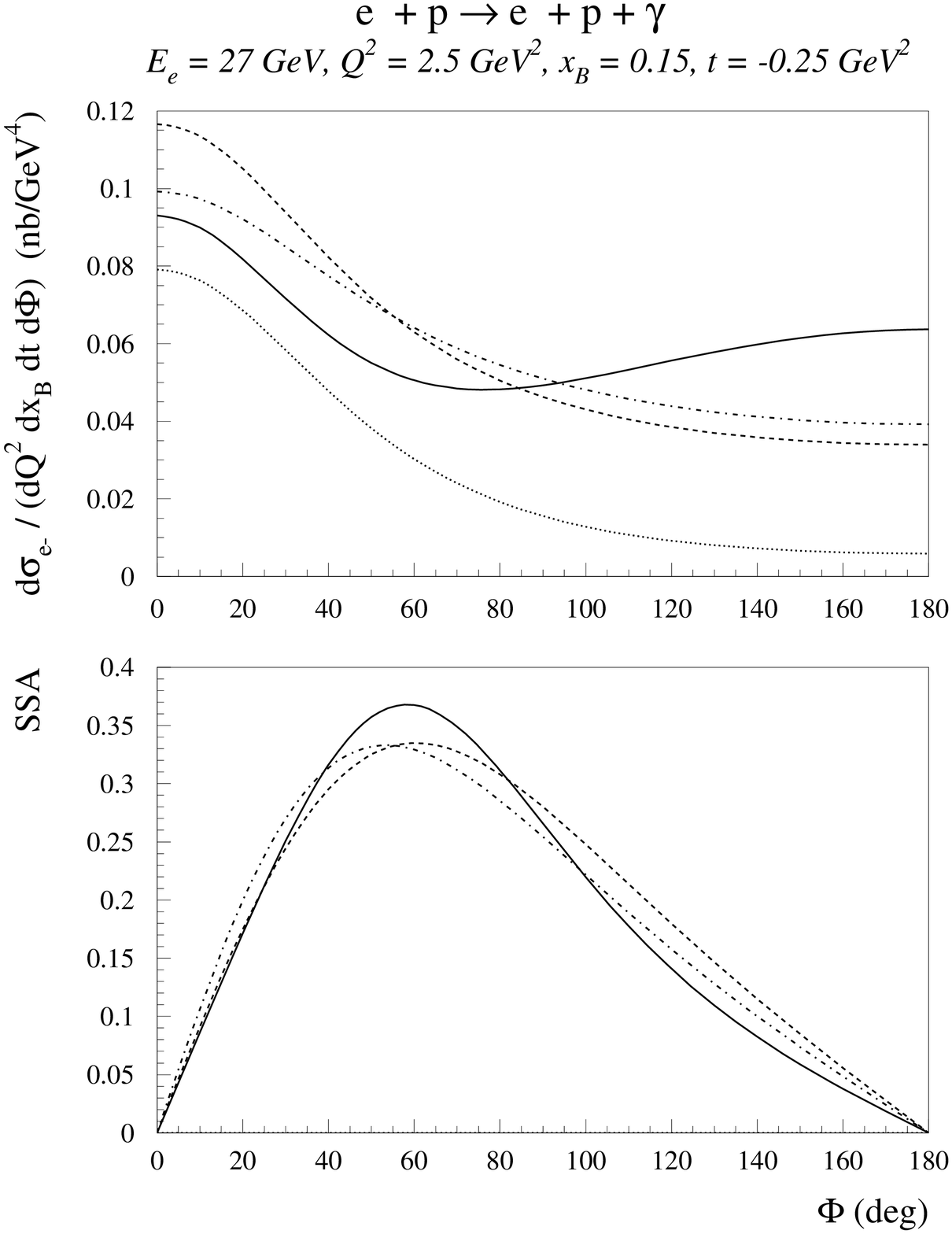}}
\caption{{\em Invariant cross section for the
$e^- p \to e^- p \gamma$ reaction (upper panel) and
DVCS SSA (lower panel) at $E_e$ = 27 GeV, for the DVCS
kinematics as indicated in the figure.
Curve conventions as in Fig.~\ref{fig:cross1}. }}
\label{fig:asymm1}
\end{figure}

\begin{figure}[h]
\epsfxsize=13cm
\centerline{\epsffile{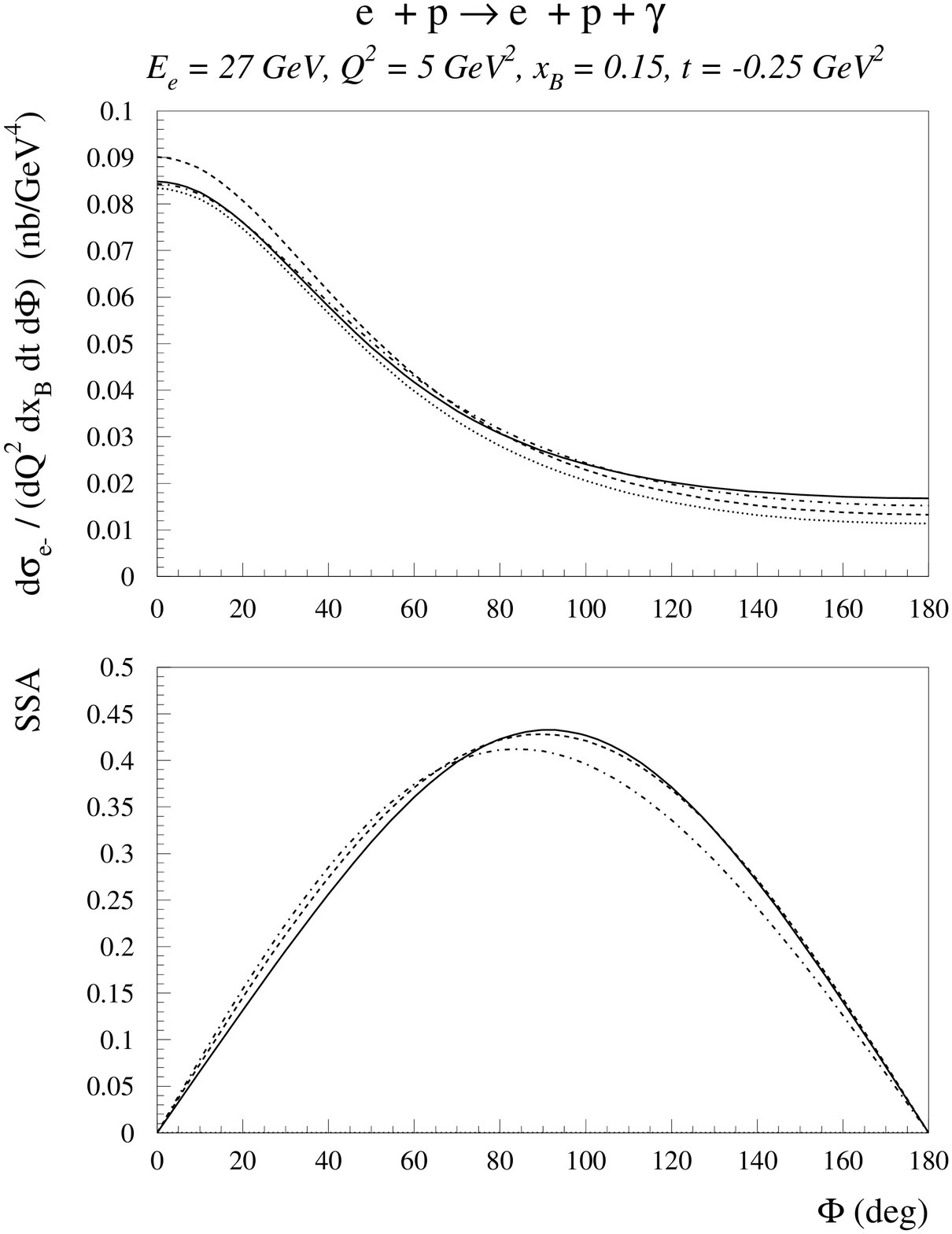}}
\caption{{\em Same as Fig.~\ref{fig:asymm1}, but for $Q^2$ = 5 GeV$^2$.}}
\label{fig:asymm2}
\end{figure}

\begin{figure}[h]
\epsfxsize=13cm
\centerline{\epsffile{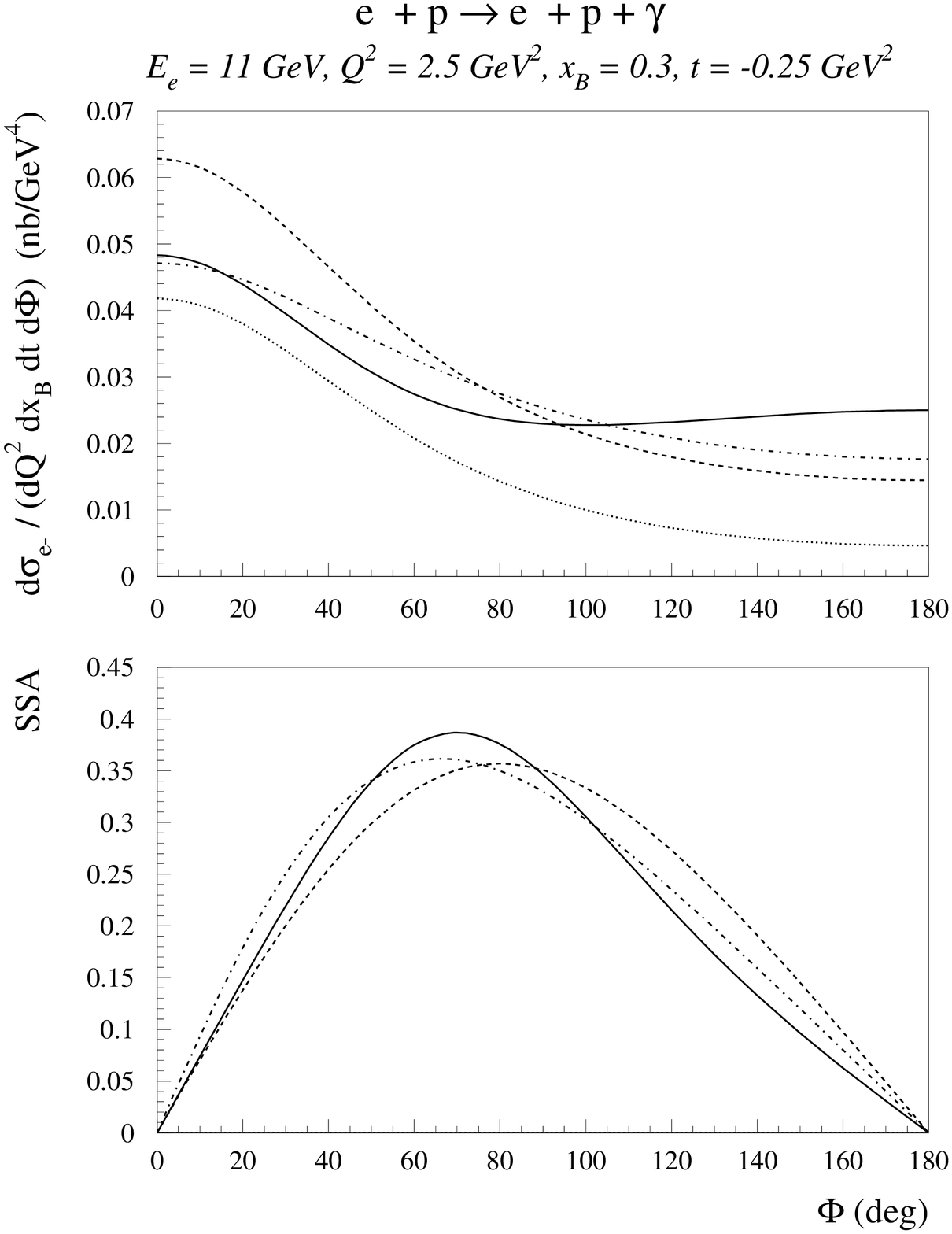}}
\caption{{\em Invariant cross section for the
$e^- p \to e^- p \gamma$ reaction (upper panel) and
DVCS SSA (lower panel) at $E_e$ = 11 GeV, for the DVCS
kinematics as indicated in the figure.
Curve conventions as in Fig.~\ref{fig:cross1}.}}
\label{fig:asymm3}
\end{figure}

In this section, we present our results for the DVCS observables.
We give all results for the invariant cross section of the
$e p \to e p \gamma$ reaction, which is differential with respect to
$Q^2$, $x_B$, $t$, and out-of-plane angle $\Phi$ ($\Phi =
0^o$ corresponds to the situation where the real photon is emitted in
the same half plane as the leptons). The invariant $e p \to e p
\gamma$ cross section is given by~:
\begin{eqnarray}
{{d \sigma} \over {d Q^2 \, d x_B \, d t \, d \Phi}} \, = \,
{1 \over {(2 \pi)^4 \, 32}} \, \cdot \, {{x_B \, y^2} \over {Q^4}}
\, \cdot \, \left( 1 + {{4 M^2 x_B^2} \over {Q^2}}\right)^{-1/2}
\, \cdot \, \biggl| T_{BH} + T_{FVCS} \biggr|^2 \, ,
\label{eq:invcs}
\end{eqnarray}
where $M$ is the nucleon mass, $y \equiv (p \cdot q) / (p \cdot k)$,
and $k$ is the initial lepton four-momentum.
In the $e p \to e p \gamma$ reaction, the final photon can be emitted
either by the proton or by the lepton.
The former process is referred to as the fully VCS process
(amplitude $T_{FVCS}$ in Eq.~(\ref{eq:invcs})), which includes
the leptonic current. The process where the photon is emitted from the
initial or final lepton is referred to as the
Bethe-Heitler (BH) process (amplitude $T_{BH}$ in
Eq.~(\ref{eq:invcs})), and can be calculated exactly.
For further technical
details of the $e p \to e p \gamma$ reaction and observables, we refer
to Ref.~\cite{GuichonVander}.
\newline
\indent
When calculating twist-3 effects in DVCS observables, we show all
results with the exact expression for the BH amplitude, i.e. we do
{\it not} expand the BH amplitude in powers 1/Q.
Also in the actual calculations, we use the exact kinematics for the
four-momenta of the participating particles
(the kinematics in Eq.~(\ref{lce}) were shown to twist-3 accuracy only
for simplicity of the presentation).
In this way we take (partially) kinematical higher twists into account.
Furthermore, when referring to the twist-2 DVCS results, we include
those higher twist effects which restore exact transversality of the
amplitude as expressed through Eq.~(\ref{transvers}). Similarly, when
referring to the twist-3 DVCS results (providing from
a longitudinally polarized virtual photon), we include the gauge
restoring higher twist terms.
\newline
\indent
In Fig.~\ref{fig:cross1}, we show the $\Phi$-dependence of the $e p
\to e p \gamma$ cross section for a lepton (either electron or
positron) of 27 GeV (accessible at HERMES).
The kinematics corresponds to the valence
region ($x_B = 0.3$) and to a ratio $t / Q^2$ = 0.1 (Remark that
increasing the ratio $t/Q^2$, increases the higher twist effects).
\newline
\indent
It is firstly seen from Fig.~\ref{fig:cross1}, that in these
kinematics, the pure twist-2 DVCS process without D-term contribution
(dashed curves) dominates the $e p \to e p \gamma$ cross section
compared to the BH process (which is sizeable only around $\Phi$ =
0$^o$, where it reaches its maximal value).
In absence of the BH, the twist-2 DVCS cross section would give a
$\Phi$-independent cross section which practically
saturates the $e p \to e p \gamma$ cross section at $\Phi$ = 180$^o$,
where the BH is vanishingly small.
The $\Phi$-dependence of the dashed curve in Fig.~\ref{fig:cross1} can
be understood as the sum of this constant twist-2 DVCS cross section,
the cross section for the BH process (whose amplitude is purely real),
and the interference of the BH with the relatively
small real part of the DVCS amplitude
(in the valence kinematics, $x_B \simeq 0.3$, shown in Fig.~\ref{fig:cross1},
the ratio of real to imaginary part of the DVCS amplitude without
D-term, is around 15 \%).
\newline
\indent
When adding to the twist-2 DVCS amplitude the purely real D-term
contribution, the resulting cross section for the full twist-2 DVCS process is
given by the dashed-dotted curves in Fig.~\ref{fig:cross1}.
At $\Phi$ = 180$^o$ (where the BH `contamination' is very small),
the predominantly imaginary DVCS amplitude
(in absence of the D-term contribution), and the
purely real D-term amplitude have only a small interference. In this
region, the DVCS cross section is enhanced by about 10 \%, when using the
chiral quark soliton model estimate of Eqs.~(\ref{dterm},\ref{eq:dterm_exp})
for the D-term.
Going to $\Phi$ = 0$^o$, the purely real D-term amplitude
interferes maximally with the BH. This interference is destructive
for the electron reaction and constructive for the positron reaction.
Therefore, the effect of the D-term can be very clearly seen in the
$\Phi$-dependence of the charge asymmetry as shown on the lower panel
of Fig.~\ref{fig:cross1}.
Including the D-term contribution, the full twist-2 DVCS charge
asymmetry changes sign and obtains a rather large value ($\approx 0.15$)
at $\Phi$ = 0$^o$. On the other hand, at  $\Phi$ = 180$^o$,
where the interference with the BH (and hence the difference between
$e^-$ and $e^+$) is small, the charge asymmetry is correspondingly small.
The pronounced $\Phi$-dependence of the charge asymmetry and its
value at $\Phi$ = 0$^o$, provides therefore a nice observable
to study the D-term contribution to the SPDs,
and to check the chiral quark soliton model estimate of
Eqs.~(\ref{dterm},\ref{eq:dterm_exp}).
\newline
\indent
Adding next the twist-3 effects in the WW approximation (full
curves), it is seen from Fig.~\ref{fig:cross1} that they induce an additional
(approximate) $\cos \Phi$ structure in the cross section.
The twist-3 effects would induce an exact $\cos \Phi$ structure only
when the BH amplitude is approximated by its leading term in an
expansion in $1/Q$, neglecting its additional $\Phi$-dependence.
In all calculations, we keep however the full
$\Phi$-dependence of the BH, due to the lepton propagators, which
complicates the interference at the lower $Q$. One sees from
Fig.~\ref{fig:cross1} that the interference of the twist-3 amplitude
with the BH + twist-2 DVCS amplitude is destructive for
$\Phi \lesssim 90^o$, and constructive for $\Phi \gtrsim 90^o$.
At $\Phi = 0^o$, the twist-3 effects reduce the full twist-2 DVCS
cross section (including the D-term) by about 25 \%, whereas at
$\Phi = 180^o$, they largely enhance the twist-2 DVCS
cross section (by about 55 \%).
\newline
\indent
To further study the twist-3 effects on the $e p \to e p \gamma$ cross
section, we show in Fig.~\ref{fig:cross2}
the cross section and charge asymmetry
at the same $E_e$, $x_B$, and $t$, as in Fig.~\ref{fig:cross1},
but at a value $Q^2$ = 5 GeV$^2$.
One firstly sees that going to higher $Q^2$, at the same $E_e$, $x_B$,
and $t$, enhances the relative contribution of the BH process compared
to the DVCS process. Consequently, the cross section follows
much more the $\Phi$-behavior of the BH process. For the twist-2
cross section, one sees again clearly the effect of the D-term which
leads, through its interference with the BH amplitude, to a charge
asymmetry of opposite sign as compared to the one for the twist-2 DVCS
process without D-term.
\newline
\indent
The twist-3 effects in the kinematics of Fig.~\ref{fig:cross2}, where
to ratio $t/Q^2$ is only half the value of Fig.~\ref{fig:cross1},
are correspondingly smaller. It is furthermore seen that the twist-3
effects induce an (approximate) structure $\sim \, A \, \cos (2 \Phi)$
in the charge asymmetry ($A \approx$ - 0.08 in the kinematics of
Fig.~\ref{fig:cross2}). Only for out-of-plane angles $\Phi \gtrsim 130^o$
one sees a deviation from the simple $\cos (2 \Phi)$ structure induced
by the twist-3 amplitude, due to the more complicated $\Phi$-dependence
when calculating the BH amplitude exactly.
\newline
\indent
In Figs.~\ref{fig:asymm1} and \ref{fig:asymm2},
we compare the $\Phi$-dependence of the DVCS cross section and of the
single spin asymmetry (SSA) corresponding to a polarized electron beam,
for the same values of $Q^2$ and $t$ as in Figs.~\ref{fig:cross1} and
\ref{fig:cross2}, but for a value of $x_B = 0.15$.
Note that, due to parity invariance, the SSA is odd in $\Phi$.
We therefore display only half of the $\Phi$ range,
i.e. $\Phi$ between 0$^o$ and 180$^o$.
\newline
\indent
When comparing Figs.~\ref{fig:cross1}, \ref{fig:cross2} and
Figs.~\ref{fig:asymm1}, \ref{fig:asymm2}, it is firstly seen that the cross
sections for the $e p \to e p \gamma$ process increases strongly when
decreasing $x_B$ at fixed $Q^2$ and fixed $t$, mainly due to the
growth of the BH amplitude.
Due to this large BH amplitude, the relative twist-3 effects in
Figs.\ref{fig:asymm1}, \ref{fig:asymm2} are smaller than the ones in
Figs.~\ref{fig:cross1}, \ref{fig:cross2}. However, the large BH
amplitude leads to a large value for the SSA through its interference
with the DVCS process. At (pure) twist-2 level, the SSA originates
from the interference of the imaginary part of the
DVCS amplitude and the (real) BH amplitude.
In case the BH is approximated by its leading term in an expansion in 1/$Q$,
the twist-2 SSA displays a pure $\sin \Phi$ structure. Due to the more
complicated $\Phi$-dependence of the BH at the lower values of $Q^2$,
this form gets distorted and its maximum displaced, as is seen by a
comparison of the twist-2 SSA in Fig.~\ref{fig:asymm1} ($Q^2$ = 2.5
GeV$^2$) and Fig.~\ref{fig:asymm2} ($Q^2$ = 5 GeV$^2$).
Note that, for practical considerations,
our ``twist-2'' DVCS calculations include kinematical higher
twist terms as well as the gauge restoring terms according to
Eq.~(\ref{transvers}). Their effect can be seen in the slight change
in the SSA (of the percent level), due to the DVCS process by itself
(i.e. when increasing the real part of the DVCS amplitude by adding
the D-term contribution, the curves for the SSAs are slightly displaced).
A more systematic treatment of target mass corrections of order
$M^2$/$Q^2$, and corrections of order $t / Q^2$ for DVCS, still remains to be
done.
\newline
\indent
We next discuss the twist-3 effects, calculated in WW approximation,
on the DVCS SSA. One sees from Fig.~\ref{fig:asymm1} that the
twist-3 corrections induce an (approximate)
$\sin (2 \Phi)$ structure in the SSA.
The amplitude of the $\sin (2 \Phi)$ term is however rather
small and the twist-3 effects change the SSA by less than 5 \% in
the kinematics corresponding to $t / Q^2 = 0.1$. It was checked that
at $x_B = 0.3$ and for a value $t / Q^2 = 0.1$,
the twist-3 effects on the SSA are of similar size.
One therefore observes that although the twist-3 effects
in WW approximation can provide a sizeable contribution
to the real part of the amplitude (Fig.~\ref{fig:cross1}),
they modify the imaginary part, and hence the SSA, to a much lesser extent.
\newline
\indent
Finally, we show in Fig.~\ref{fig:asymm3}, the corresponding cross
sections at 11 GeV (JLab), for the same $Q^2$, $x_B$ and $t$ as in
Fig.~\ref{fig:cross1}. From the cross sections, one firstly observes
again the rather large effect of the D-term,
which changes the twist-2 cross section by around 25 \%. The relative
twist-3 effects on the cross section are smaller in this case than in
Fig.~\ref{fig:cross1}, as they are on top of a larger BH amplitude.
For the SSA, it is again observed
that the twist-3 effects induce an approximate $\sin (2 \Phi)$ structure,
with amplitude of less than 5 \% in the valence kinematics considered here.

\section{\normalsize \bf Conclusions}
\label{sec5}

We have estimated the size of twist-3 effects on DVCS observables in
the Wandzura-Wilczek (WW) approximation, which allows to express the
new twist-3 SPDs in terms of twist-2 SPDs and their derivatives.
The twist-3 effects in WW approximation display therefore a new
sensitivity to the shape of the twist-2 SPDs and provide an
additional handle in the extraction of twist-2 SPDs from DVCS
observables.
The WW approximation relies on the assumption that nucleon matrix
elements of quark-gluon operators are small and can be neglected
compared to the matrix elements of quark operators. Recent data on
polarized structure functions indicate that this approximation holds to
good accuracy. Therefore, the WW approximation was considered in this
paper as a benchmark to study the size of the twist-3 effects on DVCS
observables.
It remains for future study to quantify the remaining twist-3 effects
due to matrix elements of mixed quark-gluon operators.
\newline
\indent
To provide estimates and to interpret DVCS observables, we proposed a
two-component parametrization for the SPDs of the nucleon as a sum of a
double distribution part and a D-term, so as to satisfy the
polynomiality conditions. The D-term contribution is intimately
related to the spontaneous breaking of the chiral symmetry, and was
estimated using the chiral quark soliton model. It was shown that the
effect of the D-term, which gives a purely real contribution to the
DVCS amplitude, can be very clearly seen in the dependence of the
$e p \to e p \gamma$ cross section on the out-of-plane angle $\Phi$.
Through its interference with the BH process, the D-term contribution
was seen to change the sign of the DVCS charge asymmetry, and to lead
to a large value for the charge asymmetry at $\Phi = 0^o$.
The pronounced charge asymmetry provides
therefore a clear signature to study the D-term and to check
the chiral estimate given here. 
The importance of a quantitative understanding of the D-term is
especially emphasized by the fact that it contributes considerably
to the DVCS cross section but drops out in the combination $H+E$ 
in Ji's angular momentum sum rule \cite{DVCS2}.
\newline
\indent
The twist-3 effects were seen to give a sizeable change of the real
part of the DVCS amplitude in the valence region,
and they introduce an (approximate) $\cos \Phi$ structure in the cross
section.
In kinematics where the DVCS amplitude is comparable to the
BH amplitude, one gets sizeable effects. In particular, it was shown for the
$e^- p \to e^- p \gamma$ reaction at $E_e$ = 27~GeV, in valence
kinematics ($x_B \simeq 0.3$), and for a value $t/Q^2$ = 0.1
(the value $t/Q^2$ governs kinematically the size of
higher twist effects), that the twist-3 effects reduce the full
cross section around $\Phi = 0^o$ by around 25 \%, and largely
enhance the cross section at $\Phi = 180^o$ (where the BH is very small).
On the charge asymmetry, the twist-3 effects introduce an (approximate)
$\cos (2 \Phi)$ structure, with amplitude around 10 \% in those
same kinematics.
\newline
\indent
For the single spin asymmetry, which is proportional to the imaginary
part of the DVCS amplitude, the twist-3 effects introduce an
(approximate) $\sin (2 \Phi)$ structure whose amplitude is however
rather small, so that the SSA gets changed by less than 5 \% in
typical valence kinematics for a value $t / Q^2 = 0.1$.
\newline
\indent
In summary, we found that the $e p \to e p \gamma$
cross section and charge asymmetry in the valence region,
display a sensitivity on twist-3 effects, in particular
through the real part of the amplitude.
In the WW approximation, this sensitivity to the shape of the twist-2 SPDs
may be exploited in fact in the extraction of 
those same twist-2 SPDs from DVCS observables. 
Such a systematic study of the extraction of twist-2 SPDs from
DVCS observables to twist-3 accuracy in the WW approximation,
seems to be a promising subject for future work.

\section*{\normalsize \bf Acknowledgments}
The authors like to thank K. Goeke, P.A.M. Guichon, D.~M\"uller,
and O.V.~Teryaev for useful discussions.
The work of N.K. was supported by the DFG, project No. 920585;
the work of M.V.P. was supported by DFG, BMFB and COSY;
the work of M. Vdh was supported by the DFG (SFB443).
M. Vdh also likes to thank in particular K. Goeke for
his support and warm hospitality in Bochum.

\newpage

\end{document}